\documentclass[10pt]{article}

\usepackage{amsfonts, amssymb, amsthm}
\usepackage{epsfig}
\usepackage{color}

\newtheorem{tw}{Theorem}

\newtheorem{de}{Definition}

\newtheorem{co}{Corollary}

\newtheorem{pro}{Property}

\renewcommand{\theequation}{\arabic{section}.\arabic{equation}}

\newcommand{\be}{\begin{equation}}

\newcommand{\ee}{\end{equation}}

\newcommand{\bea}{\begin{eqnarray}}

\newcommand{\eea}{\end{eqnarray}}

\usepackage{epsfig}
\usepackage{amsfonts, amssymb, amsthm}

\newcounter{orange}
\renewcommand{\theorange}{\alph{orange}}

\begin{document}

\begin{center}
{\LARGE{\bf{The eigenvalue equation for a $1$--D  Hamilton function  in deformation quantization }}}
\end{center}

\bigskip

\begin{center}
 J. Tosiek 
\end{center}

\begin{center}

{\sl  Institute of Physics, Technical University of  \L\'{o}d\'{z},\\ W\'{o}lcza\'{n}ska 219, 90-924 \L\'{o}d\'{z}, Poland.}\\
\medskip

{e-mail:  tosiek@p.lodz.pl}
\end{center}

\centerline{\today}

\begin{abstract}
The eigenvalue equation has been found for a Hamilton function in a form independent of the choice of a potential. This paper proposes 
  a modified Fedosov construction  on a flat symplectic manifold. 
 Necessary and sufficient conditions for solutions of an eigenvalue equation to be Wigner functions of pure states are presented. The $1$--D harmonic oscillator eigenvalue equation in the coordinates time and  energy is solved. A perturbation theory based on the variables time and energy is elaborated. 
\end{abstract}


\section{Introduction}

Quantum mechanics formulated in terms of linear operators acting in a Hilbert space is a potent theory in modern physics. However, although numerous problems have been solved in the framework of this mathematical model, we should be aware of its areas of weakness. The first difficulty is the lack of a  direct link between quantum mechanics and classical physics.  The second obstacle is the quantization procedure, which in its original version can only be applied to Cartesian coordinates in phase spaces of the type ${\mathbb R}^{2n}.$

It seems that deformation quantization is the  formulation of quantum mechanics,  which  
overcomes the above mentioned obstacles. Currently there exist versions of deformation quantization adapted to any symplectic or even Poisson manifold (for a review see e.g. \cite{ds}). The first version of deformation quantization, dedicated exclusively to the symplectic phase spaces  $({\mathbb R}^{2n}, \omega),$  was proposed by Moyal \cite{MO49}, who applied   the ideas of Weyl \cite{WY31}, Wigner \cite{WI32} and Groenewold \cite{GW46}.

In this paper we deal with systems with the phase space  $({\mathbb R}^{2}, \omega),$ which  can covered  with one chart. The coordinates $q$ and $p$ represent a position and a canonically conjugated momentum respectively. The symplectic form $\omega= dq \wedge dp.$ 
Observables are assumed to be smooth real functions defined in ${\mathbb R}^{2}.$ 

As a $*$--product we have chosen the {\bf Moyal product} defined by the formula (compare \cite{GW46}--\cite{ja1})
\be
\label{i1}
A *B= A \exp \left( -\frac{i \hbar}{2} \stackrel{\longleftrightarrow}{\cal P} \right) B, \;\;\; A, B \in C^{\infty}({\mathbb R}^{2}),
\ee 
where the Poisson operator $\stackrel{\longleftrightarrow}{\cal P} := \frac{\stackrel{\longleftarrow}{ \partial} }{\partial q} \frac{\stackrel{\longrightarrow}{ \partial} }{\partial p} -
\frac{\stackrel{\longleftarrow}{ \partial} }{\partial p} \frac{\stackrel{\longrightarrow}{ \partial} }{\partial q}.
 $
 We are using the sign convention compatible with Fedosov \cite{6}, \cite{7} so 
 \[
 A*B= A \cdot B -\frac{i \hbar}{2} A \stackrel{\longleftrightarrow}{\cal P} B + \ldots
 \]
 The differential definition  (\ref{i1}) of the Moyal product follows from the Fourier representation of the $*$--product \cite{bak}
 \be
 \label{i2}
 \Big(A * B \Big)(q,p)=
\frac{1}{\pi^2 \hbar^2 } \int_{{\mathbb R}^4} dq'dp'dq'' dp'' A(q',p') B(q'',p'') \exp \left[ \frac{2i}{\hbar}\Big\{
(q'-q)(p''-p)-(q''-q)(p'-p) \Big\}
\right]. 
 \ee
 The Moyal product is closed i.e.
 for all functions, for which the integrals exist
 \[
 \int_{{\mathbb R}^{2}} dqdp \, A(q,p)*B(q,p) =  \int_{{\mathbb R}^{2}} dqdp \, B(q,p)*A(q,p) =  \int_{{\mathbb R}^{2}} dqdp \, A(q,p) \cdot B(q,p) .
 \]
 By the {\bf Moyal bracket} we mean the mapping $C^{\infty}({\mathbb R}^{2}) \times C^{\infty}({\mathbb R}^{2}) \rightarrow C^{\infty}({\mathbb R}^{2})$ defined  as a factor of noncommutativity in the Moyal product
 \be
 \label{i2.1}
\{A,B\}_{M}:= \frac{1}{i \hbar} \Big(A*B -B*A \Big). 
\ee
 
 In deformation quantization
 the  counterpart of a density matrix $\hat{\varrho}$  is the quasi-probability measure in ${\mathbb R}^2$ referred to as the {\bf Wigner function} $W$. This function is related  to the density matrix by the Weyl correspondence ${\bf W}^{-1}$ (see \cite{ja1}). Therefore
 \be
 \label{i3}
W(q,p):= 
{\bf W}^{-1}\Big( \frac{1}{2 \pi \hbar}\hat{\varrho}\Big)=
\frac{1}{2 \pi \hbar}  
 \int^{+\infty}_{-\infty} d \xi \,
 \Big< q - \frac{\xi}{2} \Big| \hat{\varrho}
\Big| q + \frac{\xi}{2} \Big>  \, \mbox{exp} \left( -\frac{i\xi p}{\hbar} 
\right) . 
\ee
 In the case of a pure state, when $\hat{\varrho}=|\Psi\big>\big<\Psi|,$ the Wigner function
 equals
\be
\label{a1}
W(q,p)= \frac{1}{2 \pi \hbar} \int^{+\infty}_{-\infty} d \xi \,
\overline{\Psi} \left( q + \frac{\xi}{2} \right)
\Psi \left( q - \frac{\xi}{2} \right)  \mbox{exp} \left( -\frac{i\xi p}{\hbar} 
\right) 
\ee
or, equivalently,
\be
\label{a2}
 W(q,p)= \frac{1}{2 \pi \hbar}
 \int^{+\infty}_{-\infty} d \eta \,
\overline{\Psi}
\left( p + \frac{\eta}{2}  \right)
\Psi \left( p - \frac{\eta}{2} \right)  
 \mbox{exp} \left( \frac{i\eta q}{\hbar} \right) . 
\ee
The mean value of a function $A$ in the state represented by the Wigner function $W$ is the integral
\be
\label{a2.1}
\big< A \big> = \int_{{\mathbb R}^2} dq dp \: W(q,p) \cdot A(q,p). 
\ee 

The eigenvalue equation for a function  $A$ is of the form
\setcounter{orange}{1}
\renewcommand{\theequation} {\arabic{section}.\arabic{equation}\theorange}
\be
\label{r2.1}
A*W_{\lambda} = \lambda W_{\lambda} 
\ee
\addtocounter{orange}{1}
\addtocounter{equation}{-1}
with the additional condition 
\be
\label{r2.11}
\{A,W_{\lambda}\}_{M}=0.
\ee
\renewcommand{\theequation} {\arabic{section}.\arabic{equation}}
 By $\lambda$ we denote an eigenvalue of the function $A.$ Since we consider only a situation in which $A$ is a real smooth function defined on ${\mathbb R}^2,$ its eigenvalues are real.
The symbol $W_{\lambda}$ represents an eigenfunction of $A$ which is also a Wigner function. Hence it is real and normalizable. 
 In the literature you can also find the names: a $*$--genvalue equation for (\ref{r2.1}) and a $*$--genfunction for $W_{\lambda}$ (see \cite{ckz}). 
   
This paper  considers  the problem of finding physical solutions to Eqs. (\ref{r2.1}) and (\ref{r2.11}). We analyze the case where  observable $A$ is a $1$--D nonrelativistic  Hamilton function with a potential $V(q)$ such that the limits  $\lim_{q \rightarrow \pm \infty} V(q)= +\infty.$ The results can be generalized in a natural way to other functions and   the phase spaces ${\mathbb R}^{2n},\; n>1.$ 

There are two obstacles to be dealt with. The first one is 
the choice of a local chart, in which  the formulas (\ref{r2.1}) and (\ref{r2.11})  take a `covariant' form. We propose such a choice and present a modified Fedosov algorithm to construct the eigenvalue equation directly in this chart. The `covariant' form of the eigenvalue equation is the same  for all Hamiltonian functions. Complete information about the potential $V(q)$ is contained in symplectic connection coefficients. 
The new coordinates are the energy $H$ and the time $T$. We define this time as the time of the classical motion from a turning point but there are some alternative choices of the coordinate canonically conjugated to energy. 

 The second obstacle is a necessary and sufficient condition for a solution of (\ref{r2.1}), (\ref{r2.11}) to be a Wigner function of a pure state. We propose two criteria. 
The first one, presented in Theorem \ref{tw1} and Corollary \ref{co1}, is based on the fact that the Wigner function of a pure state is the image of a projection operator of trace $1.$ 
This condition appears in Ref. \cite{dias1}.
We write it  in both a differential form and an integral form. It seems that the latter  is more useful in calculations. 

Another criterion, presented in Theorem \ref{tw1.2}, for a solution of (\ref{r2.1}), (\ref{r2.11}) to be a Wigner function of a pure state, looks similar to the  necessary and sufficient condition for a Wigner function to represent a pure state. However, we stress that our condition can be applied to any function. This feature is extremely important for practical purposes. Indeed, an analysis, if  a given function is a Wigner function, is a complicated task. This question  is considered in \cite{dias1} and \cite{dias2}.

  Unfortunately, we were not able to write our criteria  in a covariant form. To apply them we have therefore to transform our solution of (\ref{r2.1}), (\ref{r2.11}) into the chart $(q,p)$. 

The example of the $1$--D harmonic oscillator illustrates our considerations. One can see that in this case the change of coordinates $(q,p) \rightarrow (T, H)$ radically simplifies the form of the eigenvalue equation  for the Hamiltonian.  We also present a stationary perturbation theory as another example.

We restrict our considerations to states which, in the Hilbert space formulation of quantum mechanics, are represented by normalizable vectors.   

\section{The eigenvalue equation for a Hamilton function}

In the case when the function $A$ is  Hamiltonian $H=\frac{p^2}{2m} + V(q),$ the system of equations (\ref{r2.1}), (\ref{r2.11}) is of the form (compare \cite{hms})
\setcounter{orange}{1}
\renewcommand{\theequation} {\arabic{section}.\arabic{equation}\theorange}
\be
\label{2.4a}
-\frac{p}{m} \frac{\partial W_E}{\partial q}  + \sum_{r=1,3, \ldots}^{\infty} \frac{1}{  r!} \left(\frac{i\hbar}{2}\right)^{r-1}
\frac{\partial^r V}{\partial q^r} \frac{\partial^r W_E}{\partial p^r}=0,
\ee
\addtocounter{orange}{1}
\addtocounter{equation}{-1}
\be
\label{2.4b}
\Big(\frac{p^2}{2m} + V(q) \Big) W_E -\frac{\hbar^2}{8m} \frac{\partial^2 W_E}{\partial q^2}+
\sum_{r=2,4, \ldots}^{\infty} \frac{1}{  r!} \left(\frac{i\hbar}{2}\right)^{r}
\frac{\partial^r V}{\partial q^r} \frac{\partial^r W_E}{\partial p^r}= E \, W_E,
\ee
\renewcommand{\theequation} {\arabic{section}.\arabic{equation}}
where by $E$ we denote an eigenvalue of the Hamilton function.
As may be seen, we are dealing with two partial differential equations. An explicit form of these equations depends on the smooth potential $V(q).$  The degrees of these equations are  determined by  $V(q)$ and they can be infinite.

One of methods of solving differential equations is to change the variables. We propose a special  choice of coordinates in which the   relations  (\ref{2.4a}) and (\ref{2.4b}) take a covariant form. This  covariant form highlights the geometrical nature of the eigenvalue equation.

We will transform Eqs. (\ref{2.4a}), ( \ref{2.4b}) into a system of equations, which locally looks the same for any  potential. We assume that  the potential $V(q)$ is bounded from below so we put $V(q)\geq 0$ and $\lim_{q \rightarrow \pm \infty} V(q)= \infty.$ It seems to be possible to extend our method to unbounded states as well.  


\subsection{The canonical coordinates: time and energy}

Our new canonical coordinates are: a  time $T$ and the energy $H.$ We have chosen them because for high energies when the system behaves classically, for every  state under a single constraint  $H=E= {\rm const.}$ the probability distribution is proportional to $\delta(H-E).$ Thus, also in the quantum case, the coordinate $H$ should play a dominant role. The influence of the  variable $T$ canonically conjugated to energy  vanishes in the classical limit. 

Another argument supporting this choice of coordinates has been presented in \cite{dias3}. N. C. Dias and  J. N. Prata
have shown that the phase space distribution $\delta_*(A-a)$ representing the projector $|a><a|$ on the pure eigenstate of the operator 
$\hat{A}$ for the eigenvalue $a$ is a formal $\hbar$- deformation of the generalized function $\delta(A-a).$

   We interpret the coordinate $T$  as a time of arrival. Its full description will be presented   in the next paragraph. However, formally  $T$ is a solution of the differential equation 
 \[
\frac{\partial T}{\partial q} \frac{\partial H}{\partial p} - \frac{\partial T}{\partial p} \frac{\partial H}{\partial q}= \frac{p}{m}\frac{\partial T}{\partial q} - \frac{\partial V(q)}{\partial q} \frac{\partial T}{\partial p}=1.
\]
It is defined up to a function $f(H)$  and it need not represent an actual time.
 
\begin{figure}[h]\centering
\epsfig{file=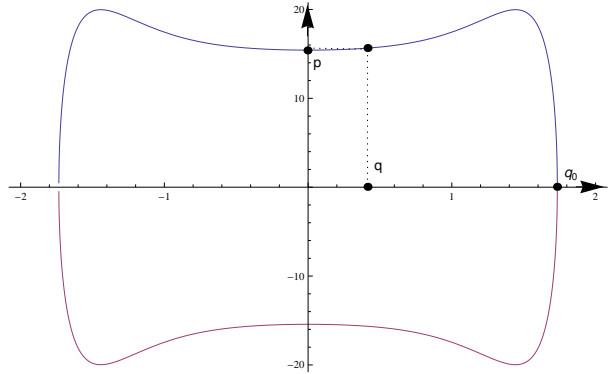,width=80mm}
\caption{The phase space diagram. }
\label{pic1.1}
\end{figure}

A canonical transformation $(q,p)\rightarrow (T, H)$ is nonsingular unless $p=0 $ and $ \frac{dV}{dq}=0.$ 
But as the measure of the set of these singular points equals $0$, they are negligible. In our model the coordinate $T$ represents the time which is necessary to reach a point $(q,p)$ on the phase space of the system from a chosen turning point $(q_0,0)$ with a fixed $H=E={\rm const.}$ Locally
\be
\label{r2.2}
T= m \int_{q_0}^q \frac{{\rm sgn}(p)}{\sqrt{2m E-2mV(z)}}dz.  
\ee
The coordinate $T$ has an intuitive geometric interpretation. Indeed, one can see that
\be
\label{r2.201}
T= \frac{\partial }{\partial E} 
\int_{q_0}^q
{\rm sgn}(p) \sqrt{2m E-2mV(z)}dz.
\ee
The integral $\int_{q_0}^q
{\rm sgn}(p) \sqrt{2m E-2mV(z)}dz$ is an area between the phase space trajectory drawn from the turning point $q_0$ to the point $q$ and the position axis (see Fig. \ref{pic1.1}). Thus  time $T$ expresses the change of this area relative to the change of energy.
\begin{figure}[h]\centering
\epsfig{file=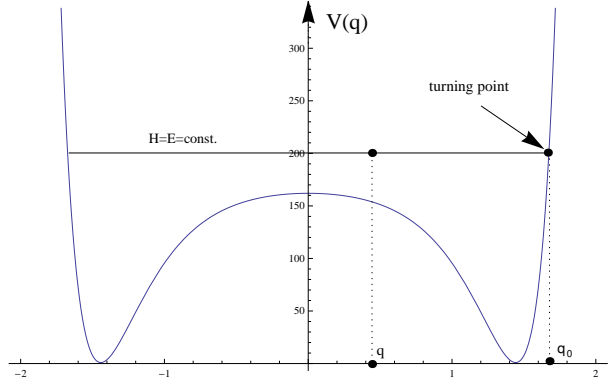,width=80mm}
\caption{A potential $V(q)$ as a function of $q$. }
\label{pic1.0}
\end{figure}

The definition of $T$ fails if $E$ equals a value of potential in its local peak, because the time of arrival to the peak is infinite. So the time for points with reverse momentum    is  not defined.

It is possible to redefine $T$ in such a way that the coordinate $T$ is finite unless $p=0 $ and $ \frac{dV}{dq}=0.$
Moreover, as the volume of a surface of constant energy equals $0,$ we have no reason  to modify formula (\ref{r2.2}).
 Finally, in the case when the potential $V(q)$ has peaks, we must cover the symplectic space $({\mathbb R}^2, \omega) $ with more than one chart.
 

\subsection{The symplectic connection}

To construct the eigenvalue equation for Hamiltonian in the new chart $(T,H)$ it is enough to transform the variables in formulas 
(\ref{2.4a}) and  (\ref{2.4b}). But this operation obscures both the geometrical character of the change and its interpretation. Therefore, we prefer another method -  writing Eqs. (\ref{2.4a}) and  (\ref{2.4b}) using a  symplectic connection. An introduction to symplectic differential geometry can be found in \cite{gel}.

\begin{de}
 The {\bf symplectic connection} $\gamma$ on a symplectic manifold $({\cal W}, \omega)$ is a torsion-free connection    satisfying  the conditions
\be
\label{d00}
\omega_{ij;k}=0,\;\;\; 1 \leq i,j,k \leq 2n= \dim {\cal W}, 
\ee
where the semicolon `$\: ;$' stands for the covariant derivative.
\end{de}

In  the Darboux coordinates the system of equations (\ref{d00}) reads
\be
\label{d1.1}
\omega_{ij;k}= - \gamma^l_{ik}\omega_{lj} - \gamma^l_{jk}\omega_{il}= \gamma_{jik}- \gamma_{ijk}=0
\ee
and
$\gamma_{ijk} :=  \gamma^l_{jk}\omega_{il}.
$
Coefficients $\gamma_{ijk}$ are symmetric with respect to the indices $\{i,j,k\}.$ 
The symplectic manifold $({\cal W}, \omega)$ endowed with the symplectic connection $\gamma$ is called a  {\bf Fedosov manifold} and it is denoted by $({\cal W}, \omega, \gamma).$

Hereafter we will work in Darboux coordinates, so locally every symplectic connection $\gamma$ will be characterized by 
 the coefficients $\gamma_{ijk}$, which are  symmetric in their indices. Local coordinates will be denoted by $(x^1, \ldots, x^{2n}).$
 
The general transformation rule for  symplectic connection coefficients is of the form \cite{my}
\be
\label{nowa1}
\gamma'_{ijk}(\tilde{x}^1, \ldots, \tilde{x}^{2n})=\frac{\partial x^l}{\partial \tilde{x}^i}\frac{\partial x^r}{\partial \tilde{x}^j}
\frac{\partial x^s}{\partial \tilde{x}^k} \gamma_{lrs}(x^1, \ldots, x^{2n}) + \omega_{rd}\frac{\partial x^r}{\partial \tilde{x}^i}
\frac{\partial^2 x^d}{\partial \tilde{x}^j \partial \tilde{x}^k}.
\ee
Locally the symplectic curvature tensor components are defined as
\be
\label{e1}
K_{ijkl}(x^1, \ldots, x^{2n}):=\omega_{iu}K^u_{jkl}=\frac{\partial \gamma_{ijl}}{\partial x^k}
- \frac{\partial \gamma_{ijk}}{\partial x^l}+ 
\omega^{st}\gamma_{til}\gamma_{sjk}
- \omega^{st}\gamma_{tik}\gamma_{sjl},
\ee
where $ \omega_{kl}\omega^{lj}=\delta^j_k.$

The phase space $({\mathbb R}^2, dq \wedge dp)$ is assumed to be symplectic flat- i.e. all  symplectic curvature tensor components (\ref{e1}) disappear. Moreover, in the coordinates $(q,p)$
all the coefficients of the symplectic connection vanish. Hence in  the new coordinates $(T,H)$ we obtain
\be
\label{nowa1.1}
\gamma_{ijk}(T,H)= \frac{\partial q}{\partial Q^i}
\frac{\partial^2 p}{\partial Q^j \partial Q^k} - \frac{\partial p}{\partial Q^i}
\frac{\partial^2 q}{\partial Q^j \partial Q^k},
\ee
where $Q^1=T, Q^2=H.$

After simple but tedious calculations we see that
\[
\gamma_{111}= -p^2 \frac{d^2 V(q)}{dq^2}- \left( \frac{d V(q)}{dq}\right)^2 \;\;,\;\;
\gamma_{112}=\frac{1}{p} \left( \frac{d V(q)}{dq} - \gamma_{111}\left( \frac{\partial T}{\partial p}\right)_q \right),
\]
\[
\gamma_{122}=-\frac{1}{p^2}\left(1+ \left( \frac{\partial T}{\partial p}\right)_q \left(\frac{d V(q)}{dq}+p \,\gamma_{112} \right) \right)\;,\; 
\gamma_{222}=\frac{1}{p^2}\left( \frac{\partial^2 T}{\partial p^2}\right)_q
+ \frac{1}{p^3} \left( \frac{\partial T}{\partial p}\right)_q
\left( 1+\left( \frac{\partial T}{\partial p}\right)_q \frac{d V(q)}{dq} -p^2 \, \gamma_{122}\right).
\]

\subsection{The Fedosov construction on a flat symplectic manifold}

Originally the Fedosov algorithm was used to introduce a Weyl type $*$--product on any symplectic manifold. Its complete description can be found in \cite{6}, \cite{7}.
In this subsection  we  present an adaptation of  the Fedosov construction of the $*$--product to the case of a symplectically flat Fedosov manifold $({\cal W},\omega, \gamma)$. This procedure allows  computing the Moyal product in an arbitrary chart without any reference to the coordinates $(q,p).$
The Einstein summation convention is used.

Let $({\cal W},\omega,\gamma)$ be a Fedosov manifold locally covered by a chart $(x^1, \ldots, x^{2n}).$
 The symbols $y^1, \ldots, y^{2n}$ represent the components of an arbitrary vector ${\bf y}$ belonging to the tangent space $T_{\tt p} {\cal W}$
 at the point ${\tt p} \in {\cal W} $ with respect to   the natural basis $\left( \frac{\partial}{\partial x^i}\right)_{\tt p}.$ 
 
We introduce the formal series of polynomials of $y^1, \ldots, y^{2n}$
\be
\label{333}
a= \sum_{z=0}^{\infty} \sum_{k=0}^{\left[\frac{z}{2}\right]} \hbar^k \tilde{a}_{k,i_1 \ldots i_{2n}}(y^{1})^{i_1} \ldots 
(y^{2n})^{i_{2n}}, 
\ee
where
$ 0 \leq i_1, \ldots, i_{2n}\leq z-2k \; , \;
i_1+ \cdots + i_{2n}= z-2k \;, \; \tilde{a}_{k,i_1 \ldots i_{2n}} \in {\mathbb C}.$   
The symbol $\left[\frac{z}{2}\right]$ denotes the integer part of $\frac{z}{2}.$ The degree of an element 
$\tilde{a}_{k,i_1 \ldots i_{2n}}$ is the sum $2k+i_1 + \ldots + i_{2n}.$ The set of the formal series (\ref{333}) at the point ${\tt p} $ is denoted as $P^*_{\tt p}{\cal W}[[\hbar]].$

In the set  $P^*_{\tt p}{\cal W}[[\hbar]]$   we define an  associative $\circ$--product.  Since this new multiplication is also ${\mathbb C}[[\hbar]]$--bilinear, it is sufficient to determine the values of the  $\circ$--product of the elements \\
$(y^{1})^{i_1} \ldots 
(y^{2n})^{i_{2n}} \circ (y^{1})^{j_1} \ldots 
(y^{2n})^{j_{2n}}.$

In Darboux coordinates (compare \cite{ja5})
\[
(y^i)^r (y^{i+n})^j \circ (y^i)^s (y^{i+n})^k
= r!\: j!\: s!\: k!\: \sum_{t=0}^{{\rm min}[r,k]+{\rm min.}[j,s]} \left( \frac{i \hbar}{2}\right)^t (y^i)^{r+s-t}(y^{i+n})^{k+j-t}
\times
\]
\setcounter{orange}{1}
\renewcommand{\theequation} {\arabic{section}.\arabic{equation}\theorange}
\be
\label{5.5}
\times \sum_{a={\rm max}[t-r,t-k,0]}^{{\rm min}[j,s,t]}(-1)^a \frac{1}{ a!\: (t-a)!\:(r-t+a)!\:(j-a)!\:(s-a)!\:(k-t+a)!},
\;\; i \leq n
\ee
\addtocounter{orange}{1}
\addtocounter{equation}{-1}
and
\be
\label{5.51}
(y^i)^r \circ (y^j)^s = (y^i)^r  (y^j)^s
\ee
\renewcommand{\theequation} {\arabic{section}.\arabic{equation}}
for $|i-j| \neq n. $

The pair $(P^*_{\tt p}{\cal W}[[\hbar]],\circ) $ is  called the  Weyl algebra. 
The sum  
$
{\cal P^*W}[[\hbar]] := \bigcup_{{\tt p} \in {\cal W}}   (P^*_{\tt p}{\cal W}[[\hbar]],\circ) 
$ is known as the Weyl algebra bundle. 

An important role in the Fedosov construction is played by 
differential forms with values in the Weyl bundle.
Locally such a $1$- form 
equals
\be
\label{9}
a= \sum_{z=0}^{\infty} \sum_{k=0}^{\left[\frac{z}{2}\right]} \hbar^k \tilde{a}_{k,i_1 \ldots i_{2n},s}
(x^1, \ldots, x^{2n})
(y^{1})^{i_1} \ldots 
(y^{2n})^{i_{2n}}
dx^{s}.
\ee
 The elements
$\tilde{a}_{k,i_1 \ldots i_{2n},s_1 \ldots s_m}
(x^1, \ldots, x^{2n}) $ are components of a smooth symmetric tensor field on ${\cal W}$ and 
$ C^{\infty}({\cal TW})\ni {\bf y}\stackrel{\rm locally }{=} y^i \frac{\partial}{\partial x^i} $ is a smooth vector field  on      
${\cal W}.$ Henceforth,
 we will omit the variables  in $\tilde{a}_{k, j_1 \ldots j_l, s_1 \ldots s_m}(x^1, \ldots, x^{2n})$. 
The differential $1$- forms   (\ref{9}) are smooth sections of the bundle  
$  {\cal P^*W}[[\hbar]] \otimes \Lambda^1 $. 

The Fedosov construction requires  the use of  
an antiderivation operator $\delta^{-1}:C^{\infty}({\cal P^*W}[[\hbar]] \otimes \Lambda^1) \rightarrow C^{\infty}({\cal
P^*W}[[\hbar]]),
$ which is defined as

\be
\label{11}
\delta^{-1} a = 
\frac{1}{l+1}\: y^k \frac{\partial }{\partial x^k}\rfloor a 
\ee
where $l$ is  the degree of $a$ in $y^j$'s and it equals the number of $y^j$'s. 
The operator $\delta^{-1}$ raises the degree of the forms from ${\cal P^*W}[[\hbar]] \otimes\Lambda^1$ in the Weyl algebra by $1$.

The exterior covariant derivative  
$
\partial_{\gamma}  a:= dx^k \wedge a_{;k}
$ of a $0$- form $a$ 
in a Darboux chart is of the form
\be
\label{14}
\partial_{\gamma} a =da + \frac{i}{ \hbar}\left( \gamma \circ a - a \circ \gamma \right).
\ee
The $1$--form   $\gamma$ of the symplectic connection  equals
$
\gamma := \frac{1}{2}\gamma_{ijk}y^iy^j dx^k.
$ 

Assume that the Fedosov manifold $({\cal W},\omega,\gamma)$ is symplectic flat.
With each  $a_0 \in C^{\infty}(\cal{W}) $ we assign
an element ${\cal P^*W}[[\hbar]] \ni a\stackrel{\rm denoted}{=} \sigma^{-1}(a_0)$ determined  by the iteration
\be
\label{20}
a= a_0 + \delta^{-1} \left( \partial_{\gamma}a \right).
\ee
Hence the component $a[z]$ of $a$ of the degree $z$ equals
\be
\label{21}
\left\{
\begin{array}{ccll}
a[0]&=&a_0,& \\
a[z]& = & \delta^{-1} \Big( \partial_{\gamma}a[z-1]\Big), & \;\;\;   z \geq 1.
\end{array}
\right.
\ee
The projection $\sigma(a)$ of $a \in {\cal P^*W}[[\hbar]] $ on the base space $\cal{W}$ is defined as 
$
 \sigma(a):=a|_{{\bf y}=0}=a_0.
$
The  $*$--product  of functions $a_0,b_0 \in C^{\infty}(\cal{W})$ calculated according to the rule
\be
\label{22}
a_0 * b_0 := \sigma \Big( \sigma^{-1}(a_0) \circ  \sigma^{-1}(b_0) \Big).
\ee
is the Moyal product written in the chart $(x^1, \ldots, x^{2n}).$

To illustrate the modified Fedosov construction we can calculate the $*$- square $L * L$ of the angular momentum  of a particle moving on the plane $XY.$ The Fedosov manifold of this particle is $({\mathbb R}^4,dx \wedge p_x + dy \wedge p_y , \gamma)$ and in the coordinates $(x,y,p_x,p_y)$ the symplectic connection $\gamma$ disappears. By definition $L= x p_y - y p_x$ and after long calculations performed in the chart  $(x,y,p_x,p_y)$ we obtain the result that the Moyal product $L * L= L \cdot L - \frac{\hbar^2}{2}.$ 

We can calculate the $*$- square $L * L$ directly in the new Darboux coordinates
\[
\left\{
\begin{array}{ccc}
r & = & \sqrt{x^2 + y^2}, \vspace{0.2cm}\\
\phi & = & \arctan \left( \frac{y}{x} \right), \vspace{0.2cm}\\
p_r & = & \frac{x}{\sqrt{x^2 + y^2}}p_x + \frac{y}{\sqrt{x^2 + y^2}}p_y, \vspace{0.2cm}\\
L & = & x p_y - y p_x.
\end{array} \right.
\]
In the new chart $(r, \phi, p_r, L)$ the symplectic form equals $\omega= dr \wedge d p_r + d \phi \wedge dL.$ The nonvanishing symplectic connection coefficients are
\[
\gamma_{112}= \frac{2L}{r^2} \; ,\; \gamma_{122}= - p_r \; ,\; \gamma_{124}= - \frac{1}{r} \; ,\; 
\gamma_{222}= - 2 L \; ,\; \gamma_{223}=r.
\]
Thus the symplectic connection $1$- form equals
\[
\gamma= \frac{L}{r^2}y^1 y^1 d \phi - \frac{1}{r}y^1 y^2 d L - p_r y^1 y^2 d \phi + \frac{2L}{r^2}y^1 y^2 d r
+ \frac{r}{2}y^2 y^2 d p_r + 
\]
\[
- \frac{p_r}{2}y^2 y^2 d r -L y^2 y^2 d \phi +r y^2 y^3 d \phi- \frac{1}{r} y^1 y^4 d \phi -\frac{1}{r} y^2 y^4 dr.
\]
From (\ref{21}) we find that
\[
\sigma^{-1}(L)= L - \frac{L}{r^2}y^1 y^1 + p_r y^1 y^2 + L y^2 y^2 -r y^2 y^3 +y^4 + \frac{1}{r} y^1 y^4.
\]
Since the product $L * L$ is defined as the projection $\sigma \Big( \sigma^{-1}(L) \circ \sigma^{-1}(L) \Big), $ we see that in fact
only three kinds of terms  contribute to the final result: $L, -r y^2 y^3 $ and $\frac{1}{r} y^1 y^4.$ The product $L \circ L = L \cdot L$ and 
\[
-r \, y^2 y^3 \circ \frac{1}{r} y^1 y^4 \stackrel{\rm (\ref{5.5})}{=} - \frac{\hbar^2}{4}+ \frac{i \hbar}{2}y^1y^3
- \frac{i \hbar}{2}y^2y^4 - y^1 y^2 y^3 y^4.
\]
But
\[
\sigma \Big(  - \frac{\hbar^2}{4}+ \frac{i \hbar}{2}y^1y^3
- \frac{i \hbar}{2}y^2y^4 - y^1 y^2 y^3 y^4 \Big)= - \frac{\hbar^2}{4}. 
\]
Thus finally $L * L = L \cdot L - \frac{\hbar^2}{2}$ as  expected.


\vspace{1cm}

Let us write  the eigenvalue equation for the Hamilton function $H= \frac{p^2}{2m}+V(q)$  
in the coordinates $(T,H).$  The symplectic connection is calculated according to the rule (\ref{nowa1.1}). 
As the Moyal product (\ref{i1}) is of the Weyl type, for any $A,B \in C^{\infty}({\mathbb R}^2)$ 
\[
A*B = \sum_{k=0}^{\infty} \hbar^k C_k(A,B),
\]
where for every $k$ the complex conjugation $\overline{C_k}(A,B)= C_k(B,A)$ and $C_k(A,B)= (-1)^k C_k(B,A).$ So
for odd $k$'s $\Re \big(C_k(A,B) \big)=0 $ and for even $k$'s $\Im \big( C_k(A,B) \big)=0. $ By $C_k$ we denote a bilinear differential operator of the order $k$. 

In our case both the Hamilton function $H$ and its Wigner eigenfunction $W_E$ are real. Therefore the requirement (\ref{r2.11}) implies 
\be
\label{jes1}
\Im (H*W_E)=\sum_{k=0}^{\infty} \hbar^{2k+1} C_{2k+1}(H,W_E)=0
\ee
and the relation  (\ref{r2.1}) reduces to
\be
\label{jes2}
\Re (H*W_E)=\sum_{k=0}^{\infty} \hbar^{2k} C_{2k}(H,W_E)=E \cdot W_E.
\ee

Applying  the Fedosov algorithm in the chart $(T,H)$ we find that  Eqs. (\ref{jes1}) and (\ref{jes2}) turn into
\setcounter{orange}{1}
\renewcommand{\theequation} {\arabic{section}.\arabic{equation}\theorange}
\be
\label{waznya}
\frac{\partial W_E}{\partial T}+ \sum_{r=3,5,\ldots} \sum_{s+t=1}^{r} \hbar^{r-1} \Theta_{r s t } \frac{\partial^{s+t} W_E }{\partial T^s \partial H^t}=0
\ee
and
\addtocounter{orange}{1}
\addtocounter{equation}{-1}
\be
\label{waznyb}
H \cdot W_E + \sum_{r=2,4,\ldots} \sum_{s+t=1}^{r} \hbar^{r} \Theta_{r s t} \frac{\partial^{s+t} W_E }{\partial T^s \partial H^t}=E W_E.
\ee
\renewcommand{\theequation} {\arabic{section}.\arabic{equation}}
A simple but tedious analysis shows that
 the coefficients $\Theta_{r s t}$ are polynomials in the symplectic connection coefficients $\gamma_{ijk}$ and their partial derivatives. The degrees of these polynomials do not exceed $r.$ 
In the coefficient $\Theta_{r s t}$  partial derivatives of the symplectic connection coefficients of the degrees less than $r-1$ appear. For example
\[
\Theta_{220}= \frac{1}{8}\gamma_{122}\;\;, \;\; \Theta_{210}= \frac{1}{8} \gamma_{111}\gamma_{222} - \frac{1}{8} \big( 
\gamma_{122}\big)^2.
\]
Eqs. (\ref{waznya}) and (\ref{waznyb}) look the same for any potential $V(q).$ The information about the potential is completely contained in the symplectic connection. The highest present power of $\hbar$ and the orders of the equations 
(\ref{waznya}) and (\ref{waznyb}) are the same as in Eqs. (\ref{2.4a}), (\ref{2.4b}).


\section{Physically acceptable solutions of an eigenvalue equation}

In this section we discuss methods of elimination of nonphysical solutions of Eqs. (\ref{r2.1}), (\ref{r2.11}).
More information about Wigner functions and their properties can be found in \cite{dias1}, \cite{dias2}, \cite{tat}--\cite{dias4}.

The formulas (\ref{a1}) and (\ref{a2}) have been written in  the coordinates $q$ and $p.$ However, since 
$W(q,p)$ is a function, we are able to deduce its properties in an arbitrary canonical atlas on the phase space ${\mathbb R}^2.$ Local coordinates in charts belonging to this atlas will be denoted by $Q$ and $P.$ Below are listed several properties of the Wigner functions of pure states. 
\begin{pro}
\label{pro1}
A function $ W(Q,P)$ of a pure state,  as a Wigner function, is real i.e. $W(Q,P)= \overline{W}(Q,P).$
\end{pro}
\begin{pro}
The integral
\be 
\label{a2.5}
\int_{{\mathbb R}^2} dQdP \: W(Q,P)=1.
\ee
\end{pro}
Moreover, since $ W(Q,P)$ represents a pure state and the Moyal product is closed, the following equality holds.
\begin{pro}
 \be
\label{a3}
\int_{{\mathbb R}^2} dQdP \: W^2(Q,P)= \frac{1}{2\pi \hbar}.
\ee
\end{pro}
In a case where the symplectic space $({\mathbb R}^2, \omega)$ is covered by more than one chart,  the integral
$\int_{{\mathbb R}^2}  dQdP f(Q,P)$ represents the number $\int_{{\mathbb R}^2} \omega \cdot f  .$

As  is known from the Schwarz inequality, we can estimate the function $W(Q,P).$ Indeed, 
\begin{pro}
\be
\label{a4}
|W(Q,P)| \leq \frac{1}{\pi \hbar}.
\ee
\end{pro}
This inequality implies that every Wigner function depends on $\hbar.$
The Dirac constant is a real positive parameter. From the mathematical point of view we can choose its value arbitrarily. If the function $W(Q,P)$ did not depend on $\hbar,$  the limit $\hbar \rightarrow \infty$ of (\ref{a4}) would result in  $W(Q,P)=0.$

\begin{pro}
\label{pro5}
The Wigner function $W(q,p)$ of a pure state is a continuous function 
with respect to $q$ for any $p$ and a continuous function with respect to $p$ for any $q$ 
on ${\mathbb R}^2.$ 
\end{pro}
\underline{Proof}

Let us consider the function 
\[
F(q)=\int^{+\infty}_{-\infty} d \xi \,
\overline{\Psi} \left( q + \frac{\xi}{2} \right)
\Psi \left( q - \frac{\xi}{2} \right)   , 
\]
 where $\Psi \in L^2({\mathbb R}).$ This function is well defined for every $q \in {\mathbb R}.$ Indeed, changing the variables in the integral we can write
 \[
 F(q)= 2 \int^{+\infty}_{-\infty} du \, \overline{\Psi}(u) \Psi (2q-u).
 \]
 Both functions $\Psi(u)$ and $\Psi (2q-u)$ are elements of $L^2({\mathbb R})$ for every $q$ so $F(q)$ is their scalar product.
 Hence $F(q)$ is well defined and  finite for every real $q$.
 
 Moreover, 
 \[
 \tilde{F}(q,p)=\frac{1}{2 \pi \hbar} \int^{+\infty}_{-\infty} d \xi \,
\overline{\Psi} \left( q + \frac{\xi}{2} \right)
\Psi \left( q - \frac{\xi}{2} \right)  \mbox{exp} \left( -\frac{i\xi p}{\hbar} 
\right) 
 \]
 as the Fourier transform of the function $\overline{\Psi} \left( q + \frac{\xi}{2} \right)
\Psi \left( q - \frac{\xi}{2} \right)$  is a continuous function of $p$ for every $q$ (compare \cite{schwartz}).
Repeating the same operations for the formula (\ref{a2}) we conclude that the Wigner function $W(q,p)$ of a pure state is defined and  finite at every point $(q,p).$ Moreover,  it is continuous with respect to $q$ for every $p.$ \rule{2mm}{2mm}

Properties \ref{pro1}--\ref{pro5} are necessary but not sufficient conditions for solutions of the eigenvalue equation (\ref{r2.1}), (\ref{r2.11}) to be Wigner functions. 

Now let us  consider the necessary and sufficient conditions for a function $W(Q,P)$ to be a Wigner function of a pure state. 
\begin{tw} \cite{dias1}
\label{tw1}
A real function $W(Q,P)$ defined on the phase space ${\mathbb R}^2$ is a Wigner function on a pure state if and only if 
\begin{enumerate}
\item
 $\int_{{\mathbb R}^2}  dQdP \, W(Q,P)=1$ and 
 \item
$
W(Q,P)*W(Q,P)= \frac{1}{2\pi \hbar}W(Q,P).
$
\end{enumerate}
\end{tw}
The proof of this statement is a straightforward consequence of the fact that 
in the Hilbert space formulation of quantum mechanics pure states are represented by projection operators of  trace $1.$
By definition, the projection operator is an operator which is  self-adjoint and idempotent.
Since the Weyl correspondence ${\bf W}^{-1}$ establishes a one-to-one relation between  operators and functions in the space ${\mathbb R}^2,$ 
  we see that Theorem \ref{tw1} holds.

Unfortunately, the function  $W(Q,P)$ usually contains arbitrary great negative powers of $\hbar.$  Therefore calculation of the Moyal product $W(Q,P)*W(Q,P)$ using the differential formula (\ref{i1}) can be extremely difficult. However, in  the coordinates $(q,p)$ we can apply the integral definition (\ref{i2}) of the $*$--product. Hence
\begin{co}
\label{co1}
A   necessary and sufficient condition for a real function $W(Q,P)$ to represent a pure quantum state is that 
\begin{enumerate}
\item
$\int_{{\mathbb R}^2} dQdP \, W(Q,P)=1$ and
\item
\be
\label{a5}
\frac{2}{\pi \hbar } \int_{{\mathbb R}^4} dq'dp'dq'' dp'' W(q',p') W(q'',p'') \exp \left[ \frac{2i}{\hbar}\Big\{
(q'-q)(p''-p)-(q''-q)(p'-p) \Big\}
\right]= W(q,p).
\ee
\end{enumerate}
\end{co}
 The main disadvantage of Corollary \ref{co1} is the fact  it refers to the  canonical coordinates $q$ an $p.$ We can write formula (\ref{a5}) in an arbitrary chart but to do this we have to substitute the original  variables $q,q',q'',p,p',p''$ for  new ones: $Q,Q',Q'',P,P',P''.$
 
 For every real function $W(q,p)$, for which the integral  (\ref{a5}) exists, the imaginary part of (\ref{a5}) disappears. Therefore the condition (\ref{a5}) is equivalent to 
 \be
\label{a6}
\frac{2}{\pi \hbar } \int_{{\mathbb R}^4} dq'dp'dq'' dp'' W(q',p') W(q'',p'') \cos\left[ \frac{2}{\hbar}\Big\{
(q'-q)(p''-p)-(q''-q)(p'-p) \Big\}
\right]= W(q,p).
\ee
The argument of the function $\cos$ can be expressed in terms of vector products. Indeed, let $\vec{V}=(q,p,0), \vec{V}'=(q',p',0),
\vec{V}''=(q'',p'',0), \vec{e}=(0,0,1).$ The relation (\ref{a6}) becomes
\be
\label{a6.1}
\frac{2}{\pi \hbar } \int_{{\mathbb R}^4} \,\omega' \,\omega'' \,W(\vec{V}') \,W(\vec{V}'')\, \cos\left[ \frac{2}{\hbar} \vec{e} \cdot \Big\{
\vec{V}' \times \vec{V}'' + \vec{V} \times \vec{V}' + \vec{V}'' \times \vec{V}
 \Big\}
\right]= W(\vec{V}).
\ee 
The condition (\ref{a6.1}) can be used in any chart preserving the linear structure of ${\mathbb R}^2.$
 
 Another version of the sufficient and necessary condition for a function defined on ${\mathbb R}^2$ to be a Wigner function of a pure state is formulated below. It should be emphasized that it works only in  the coordinates $(q,p)$. It looks similar to the well known necessary and sufficient condition for a Wigner function to represent a pure state \cite{tat}.
However,  contrary to the aforementioned result,  it applies to an {\bf arbitrary} function. 
 \begin{tw}
 \label{tw1.2}
 A real  function $W(q,p)$ defined on the phase space ${\mathbb R}^2$ is a Wigner function of a pure state if and only if
 \begin{enumerate}
 \item
 \label{a60}
 at every point $(q,p) \in {\mathbb R}^2$  is continuous with respect to $q$ and with respect to $p$,
 \item
 \label{a61}
 $\int_{{\mathbb R}^2} dqdp W(q,p)=1,$
 \item
 \label{a63}
 for every $q_1, q_2 \in {\mathbb R}$  there is 
 \[
 \varrho(q_1,q_2)= f(q_1)g(q_2),
 \]
 where 
 \[
 \varrho(q_1,q_2) := \int_{{\mathbb R}} dp \,W \left(\frac{q_1+q_2}{2},p \right)\exp \left[ \frac{ip(q_1-q_2)}{\hbar}\right]. 
 \]
 \end{enumerate}
 \end{tw}
 For a differentiable function $\varrho(q_1,q_2)\neq 0$ the condition (\ref{a63} ) is usually written in the form
 \[
 \frac{\partial^2 \ln \varrho(q_1,q_2)}{\partial q_1 \partial q_2}=0.
 \]
 \underline{Proof}
 \newline
 "$\Rightarrow$"
 The function $W(q,p)$ is a Wigner function of a pure state. Hence it is real, continuous with respect to $q$ and $p$ and it satisfies  the condition (\ref{a61} ).
 As  is well known \cite{tat}, every Wigner function of a pure state fulfills (\ref{a63} ).
\newline
"$\Leftarrow$"
The function $\varrho$ can be written in another form 
\[
\varrho(x,y)= \int_{{\mathbb R}} dp \,W \left(x,p \right)\exp \left[ \frac{ipy}{\hbar}\right].
\]
From the assumption (\ref{a60} ) we notice that
for every fixed $y$ the function  $W \left(x,p \right)\exp \left[ \frac{ipy}{\hbar}\right]$ is a continuous function of $p.$ Therefore the integral $\int_{{\mathbb R}} dp \,W \left(x,p \right)\exp \left[ \frac{ipy}{\hbar}\right]$
is a continuous function of $x$ (see \cite{schwartz}). Moreover, this integral is, up to a factor,  the inverse Fourier transform of $W \left(x,p \right).$ This observation implies that for every $x$ the function $\varrho(x,y)$ is a continuous function of $y.$

The relation between coordinates $(x,y)$ and $(q_1,q_2)$ is linear. Thus $\varrho(q_1,q_2)$ is well defined at every point $(q_1,q_2)$.

Since the function  $W(q,p)$ is real, there must be $\overline{\varrho}(q_1,q_2)= \varrho(q_2,q_1).$ Therefore from (\ref{a63} )
\[
\overline{f}(q_1)\overline{g}(q_2)=f(q_2)g(q_1).
\]
At all points $(q_1,q_2)$ at which $\varrho(q_1,q_2) \neq 0,$ there must be 
$\frac{\overline{f}(q_1)}{g(q_1)}= \frac{f(q_2)}{\overline{g}(q_2)}= A \neq 0,$ where $A$ is a constant. In the case $q_1=q_2,$ we obtain
$\frac{\overline{f}(q_1)}{g(q_1)}= \frac{f(q_1)}{\overline{g}(q_1)}= A$ so $A$ is a real number.
Finally 
\be
\label{a5a}
|f(q)|^2=A^2 |g(q)|^2.  
\ee
At all  points at which $\varrho(q_1,q_2)=0,$ we put $f(q_1)=g(q_2)=0.$ This means that if $\varrho(q,q)=0,$ then the formula (\ref{a5a}) also remains true. Hence the equality (\ref{a5a}) holds for every $q \in {\mathbb R}.$

Let us define a new function $F(q):= \frac{1}{\sqrt{|A|}}f(q).$ So
\be
\label{a7}
\varrho(q_1,q_2)= \sqrt{|A|}\, F(q_1) \cdot \sqrt{|A|} \:\overline{F}(q_2) \,\frac{1}{A}= {\rm sgn} (A) F(q_1)\overline{F}(q_2).
\ee
From the definition of function $\varrho(q_1,q_2)$  we can see that
\be
\label{a7.1}
\varrho(q_0,q_0)= \int_{\mathbb R} dp W(q_0,p) \stackrel{\rm (\ref{a7})}{=} {\rm sgn} (A) |F(q_0)|^2.
\ee
 Assume that $A<0.$ Then for every $q_0 \in {\mathbb R} $ the integral $\int_{\mathbb R} dp W(q_0,p)<0. $ Therefore
 $\int_{{\mathbb R}^2} dqdp W(q,p)<0 $ which is impossible according to assumption (\ref{a61} ). Hence $A>0.$ 
 
 As $\int_{\mathbb R} dq_0 \int_{\mathbb R} dp W(q_0,p)=1,$ we see
from (\ref{a7.1}) we see that $|F(q_0)|^2$ can be interpreted as a spatial probability density.
 
 A solution of (\ref{a7.1}) with respect to $F(q_0)$  is $F(q_0)= \sqrt{\int_{\mathbb R} dp W(q_0,p)} \:\exp(i \phi).$ The point $q_0$ can always be chosen in such a way that $F(q_0) \neq 0.$ 
 
 We put
 \[
 F(q):= \frac{\varrho(q,q_0)}{\overline{F}(q_0)}= \frac{1}{\sqrt{\int_{\mathbb R} dp W(q_0,p)}} \:\exp( i \phi)
 \cdot \int_{{\mathbb R}} dp \,W \left(p,\frac{q+q_0}{2} \right)\exp \left[ \frac{ip(q-q_0)}{\hbar}\right].
 \]
 The function $F(q)$ belongs to space $ L^2({\mathbb R}),$ because
\[
\int_{{\mathbb R}^2} dqdp \, W(q,p)= \int_{\mathbb R} dq\, |F(q)|^2 =1
\]
 so it represents a pure quantum state. \hspace{0.2cm} \rule{2mm}{2mm}
 
 Another sufficient and necessary condition for a square integrable function to be a pure state Wigner function was presented in \cite{dias1}. It would  also be possible to apply the Narcowich- Wigner spectrum. A detailed presentation of this topic can be found in \cite{dias2}. However, it should be remembered that exclusive analysis of  the Narcowich- Wigner spectrum
  is insufficient to identify non-Gaussian pure states.
 
\section{The examples}

Probably the most spectacular example of an application of the method proposed in our paper is the $1$--D harmonic oscillator. We presented a construction of the eigenvalue equation based on a change of canonical coordinates for the $1$--D oscillator  in \cite{my}. Here we recall briefly  the main steps of this procedure  and  concentrate on a  selection of physically acceptable solutions.

 The Hamilton function of the harmonic oscillator in  the coordinates $(q,p)$ is of the form
 \be
 \label{a8}
H= \frac{p^2}{2}+\frac{q^2}{2}. 
\ee
We put the mass $m=1$ and the frequency $\omega=1.$

\begin{figure}[h]\centering
\epsfig{file=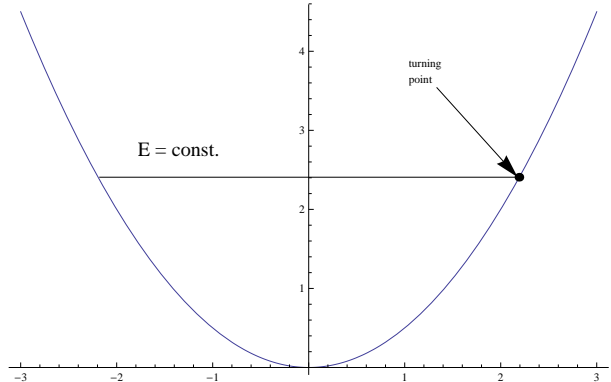,width=80mm} 
\caption{The harmonic potential}
\label{pic1}
\end{figure}

Applying the relation (\ref{r2.2}) to the potential illustrated on  Fig. \ref{pic1} we see that
\be
\label{a9}
\left\{ 
\begin{array}{cccl}
q & =&  \sqrt{2H} \cos T &\\
p & = & - \sqrt{2H} \sin T,& \;\; 0 < H \;,\; 0 \leq T < 2 \pi.
\end{array}
\right. 
\ee 

This transformation is singular at the point $(q=0,p=0)$ but this is the set of the measure $0.$ 
It is assumed that in the coordinates $(q,p)$ the symplectic connection $\gamma$ disappears. 

From the transformation rule (\ref{nowa1.1}) we find that in the coordinates $(T, H)$ the  symplectic connection coefficients are
\[
\gamma_{111}= - 2H \;\; , \;\; \gamma_{112}=0 \;\; , \;\;
\gamma_{122}=- \frac{1}{2 H} \;\; , \;\; \gamma_{222}=0.
\] 
Hence the symplectic connection $1$--form (\ref{14}) equals
\[
\gamma= -  H y^1 y^1 d T - \frac{1}{2 H }y^1 y^2 d H - \frac{1}{4 H } y^2 y^2 d T.
\]
Applying the Fedosov algorithm we find that the eigenvalue equation for the Hamiltonian is of the form
\setcounter{orange}{1}
\renewcommand{\theequation} {\arabic{section}.\arabic{equation}\theorange}
\be
\label{a10a}
\Big(H-E \Big) W_E+ \frac{i \hbar}{2}\frac{\partial W_E}{\partial T}- \hbar^2
\left(\frac{H}{4}\frac{\partial^2 W_E}{\partial H^2}+ \frac{1}{4} \frac{\partial W_E}{\partial H} 
+ \frac{1}{16 H} \frac{\partial^2 W_E}{\partial T^2}
\right)=0,
\ee
\addtocounter{orange}{1}
\addtocounter{equation}{-1}
\be
\label{a10b}
i \hbar\frac{\partial W_E}{\partial T}=0.
\ee
\renewcommand{\theequation} {\arabic{section}.\arabic{equation}}
From the second equality we notice that $\frac{\partial W_E}{\partial T}=0.$ Thus the eigenvalue equation (\ref{a10a}) reduces to the ordinary differential equation
\be
\label{a11}
\Big(H-E \Big) W_E- \hbar^2
\left(\frac{H}{4}\frac{d^2 W_E}{d H^2}+ \frac{1}{4} \frac{d W_E}{d H} 
\right)=0.
\ee
The function $W_E(H)$ can be written in the form $W_E(H)= \exp \left( -\frac{2H}{\hbar} \right)\, w_E(H).$ By an easy substitution we can see that the function $w_E(H)$ is a solution of the equation
\[
H \frac{d^2 w_E}{d H^2} + \left(1-\frac{4H}{\hbar} \right)\frac{d w_E}{d H} - \frac{2 \hbar - 4E}{\hbar^2}w_E=0.
\]
Introducing the new variable $y= \frac{4H}{\hbar}$ we can conclude that $w_E(y)$ is a solution of the differential equation
\be
\label{a12}
y \frac{d^2 w_E}{d y^2} + \left(1- y  \right)\frac{d w_E}{d y} - \frac{ \hbar - 2E}{2 \hbar}w_E=0.
\ee
As may be seen   in \cite{lan}, its solution is the linear combination
\[
w_E= C_1 y^{\frac{2E-\hbar}{2 \hbar}}G\left( \frac{\hbar- 2E}{2 \hbar}, \frac{\hbar- 2E}{2 \hbar}, -y \right)+
C_2 e^y y^{- \frac{\hbar + 2E}{2 \hbar}} \, G \left( \frac{\hbar + 2E}{2 \hbar}, \frac{\hbar + 2E}{2 \hbar}, y\right). 
\]
Symbols $C_1, C_2$ denote real numbers and the function
\[
G(\alpha, \beta, y):= 1 + \frac{\alpha \beta}{1! \, y }+ \frac{\alpha (\alpha +1)\beta (\beta +1)}{2! \, y^2 } + \ldots
\]
Hence  the solution of (\ref{a11}) is
\[
W_E(H)= C_1 \exp \left( -\frac{2H}{\hbar} \right) \left( \frac{4H}{\hbar} \right)^{- \frac{\hbar- 2E}{2 \hbar}}
G\left( \frac{\hbar- 2E}{2 \hbar}, \frac{\hbar- 2E}{2 \hbar}, - \frac{4H}{\hbar} \right)+
\]
\be
\label{a13} 
+ C_2 \exp \left( \frac{2H}{\hbar} \right) \left( \frac{4H}{\hbar} \right)^{- \frac{\hbar+ 2E}{2 \hbar}}
G\left( \frac{\hbar+ 2E}{2 \hbar}, \frac{\hbar+ 2E}{2 \hbar},  \frac{4H}{\hbar} \right). 
\ee
We know that
eigenvalues of the Hamilton function $H$ satisfy the inequality $ E \geq 0.$

The integral $\int_{0}^{\infty} d H \exp \left( \frac{2H}{\hbar} \right) \left( \frac{4H}{\hbar} \right)^{- \frac{\hbar+ 2E}{2 \hbar}}
G\left( \frac{\hbar+ 2E}{2 \hbar}, \frac{\hbar+ 2E}{2 \hbar},  \frac{4H}{\hbar} \right)$ is divergent.

Moreover, unless $-\alpha \in {\cal N },$  the  series $G(\alpha, \alpha, y)$ is divergent for any finite $y.$ So there must be $E= \hbar(n+ \frac{1}{2}), \; n \in {\cal N}.$ But $\frac{ (-1)^n }{n!} y^n G(-n,-n,-y)= L_n(y),$ where the symbol $L_n(y)$ denotes the Laguerre polynomial defined as $L_n(y):= \sum_{m=0}^n (-1)^m 
\left( \begin{array}{c}
n \\n-m
\end{array}
\right) \frac{y^m}{m!}.
$ 

We can see  that the  solutions of the eigenvalue equation of the harmonic oscillator belong to the set of functions
\be
\label{a14}
W_n(H)= C_n \exp \left( -\frac{2H}{\hbar} \right) L_n \left( \frac{4H}{\hbar} \right), \;\; n\in {\cal N}.
\ee
At this moment, we do not know whether all values of $n \in {\cal N}$  are physically acceptable. To answer this question we apply Theorem \ref{tw1.2}.

Every function $W_n(H)$ is real and continuous. They are also normalizable so for $C_n= \frac{(-1)^n}{\pi \hbar}$ (compare \cite{ryz}) the assumption 
 (\ref{a61} ) is fulfilled. We therefore concentrate on the integral
\[
\int_{\mathbb R} dp \, \frac{(-1)^n}{\pi \hbar} \exp \left( - \frac{4 p^2  + (q_1+q_2)^2}{4 \hbar}\right) L_n\left( \frac{4 p^2  + (q_1+q_2)^2}{2 \hbar}\right) \exp \left( \frac{i p (q_1 - q_2)}{\hbar}\right).
\]
As  can be verified using  the computer  program Mathematica, this integral equals
\[
 \frac{1}{2^n \, n!\,\sqrt{\pi \hbar}} \exp\left( -\frac{q_1^2}{2 \hbar}\right)H_n\left( \frac{q_1}{\sqrt{\hbar}}\right) \cdot \exp\left( -\frac{q_2^2}{2 \hbar}\right)H_n\left( \frac{q_2}{\sqrt{\hbar}}\right).
\]
It may be  concluded that the function $\varrho(q_1,q_2)$ is a product of two functions depending only on the variables $q_1, q_2$ respectively. From Theorem \ref{tw1.2} every function of the form (\ref{a14}) is a physically acceptable  eigenvalue function. This result is in agreement with the solution  obtained from the Schroedinger  equation
\be
\label{ost}
- \frac{\hbar^2}{2} \frac{d^2}{dq^2} \Psi_E(q) + \frac{q^2}{2} \Psi_E(q)= E \Psi_E(q)
\ee
via the relation (\ref{a1}). Indeed, as  is well known \cite{lan}, the solutions of (\ref{ost}) are of the form 
$\Psi_E(q)= \sqrt{\frac{1}{2^n n!\, \sqrt{\pi \hbar}} }\exp\left( -\frac{q^2}{2 \hbar}\right)H_n\left( \frac{q}{\sqrt{\hbar}}\right)$ for the eigenvalues
$E= \hbar(n+ \frac{1}{2}), n=0,1, \ldots. $

\vspace{1cm}
The second example is  a $1$st order stationary perturbation theory. Consider a given Hamiltonian
\[
H=H_0 + \lambda H_1(T_0,H_0),
\]
where $\lambda \in {\mathbb R}.$ Focusing on an area of the phase space $({\mathbb R}^2, dT_0 \wedge dH_0 ),$ in which $|\lambda H_1(T_0,H_0)| \ll |H_0|,$ the new `perturbed' variable is
\[
T= T_0 + \lambda T_1(T_0,H_0) + \ldots.
\] 
As the variables $(T,H)$ are expected to be canonical, in the linear approximation with respect to $\lambda$ we find that the function $T_1(H_0,T_0)$ must satisfy the partial differential equation
\[
\frac{\partial T_1(T_0,H_0)}{\partial T_0}+ \frac{\partial H_1(T_0,H_0)}{\partial H_0}=0.
\]
In the $1$st order approximation we deal with the canonical coordinates
\be
\label{mar1}
\left\{
\begin{array}{ccc}
T& = & T_0 + \lambda T_1(H_0,T_0)
, \\
H& = & H_0 + \lambda H_1(H_0,T_0).
\end{array} 
\right.
\ee
From (\ref{mar1}), up to linear terms in $\lambda$ we have 
\be
\label{mar2}
\left\{
\begin{array}{ccc}
T_0& = & T - \lambda T_1(T,H), \\
H_0& = & H - \lambda H_1(T,H).
\end{array} 
\right.
\ee
Further considerations will be addressed in the  linear approximation with respect to $\lambda.$

In the coordinates $(T_0,H_0)$ a flat symplectic connection was determined by the coefficients $\gamma_{111},
 \, \gamma_{112}, \, \gamma_{122},$ and $\gamma_{222}.$ During the transformation of the coordinates $(T_0,H_0) \longrightarrow (T,H)$ these connection coefficients change according to the rule   (\ref{nowa1}). For example
\[
\gamma'_{111}(T,H)= \gamma_{111}(T,H)- \lambda \left( \frac{\partial \gamma_{111}(x,y)}{\partial x}\Big|_{(T,H)}T_1(T,H)
+ \frac{\partial \gamma_{111}(x,y)}{\partial y}\Big|_{(T,H)}H_1(T,H) 
+ \right.
\]
\[ 
\left.
+3 \gamma_{111}(T,H) \frac{\partial T_1(x,y)}{\partial x}\Big|_{(T,H)}
+ \gamma_{112}(T,H) \frac{\partial H_1(x,y)}{\partial x}\Big|_{(T,H)} 
+ \frac{\partial^2 H_1(x,y)}{\partial x^2}\Big|_{(T,H)} 
\right).
\]
Thus the new symplectic connection $1$- form is
\[
\gamma'(T,H)= \gamma(T,H) + \lambda \gamma_1(T,H)
\]
and
 the new $\tilde{*}$- product of functions $A$ and $B$ reads
\be
\label{mar3}
A \tilde{*} B = A * B + \lambda \, A \overline{*}B. 
\ee
The second term $\lambda \, A \overline{*}B$ represents a `correction' to the undisturbed product $A*B.$

 We intend to solve the eigenvalue equation
 \be
 \label{mar4}
H  \, \tilde{*} \, W_E = E \, W_E.
\ee
To this end 
we represent the  eigenvalue $E$ as the series
\[
E= E_0 + \lambda E_1 + \ldots
\] 
and its eigenfunction as
\[
W_E= W_{E0} + \lambda W_{E1} + \ldots 
\]
Inserting these series into (\ref{mar4}) we obtain the system of equations
\setcounter{orange}{1}
\renewcommand{\theequation} {\arabic{section}.\arabic{equation}\theorange}
\be
\label{mar4.5}
H * W_{E0}= E_0 \, W_{E0},
\ee
\addtocounter{orange}{1}
\addtocounter{equation}{-1}
\be
\label{mar5}
H * W_{E1} + H \, \overline{*} \, W_{E0}= E_1 W_{E0} + E_0 W_{E1},
\ee
\[
\vdots
\]
\renewcommand{\theequation} {\arabic{section}.\arabic{equation}}

Let us assume  that the general solution of Eq. (\ref{mar4.5}) has been found  and  its physically acceptable part  $W_{E0}$ has been extracted. 
The relation (\ref{mar5}) can be written as
\be
\label{mar6}
(H-E_0) *W_{E1}= E_1 W_{E0}- H \, \overline{*} \, W_{E0}.
\ee
Multiplying both sides of (\ref{mar6}) by $W_{E0}*$ on the left-hand side we obtain
\[
0= \frac{1}{2 \pi \hbar} E_1 \, W_{E0} - W_{E0}* \big( H \, \overline{*} \, W_{E0} \big)
\]
The first correction to the energy is
\[
E_1= 2 \pi \hbar \int_{{\mathbb R}^2} dT dH\, W_{E0}(T,H) \big( H \, \overline{*} \, W_{E0} \big) (T,H).
\]
Now, since (\ref{mar6}) is a linear nonhomogeneous equation with respect to $W_{E1}$ and we have already found the general solution to  the homogeneous equation (\ref{mar4.5}), we are able to derive the function $W_{E1}.$


 \section{Conclusions}
 
 We have proposed a method of solving the eigenvalue equation for a $1$--D Hamiltonian based on a change of canonical coordinates. Instead of the coordinates $(q,p)$ we apply the coordinates  time $T$ and the Hamilton function $H.$ This change of a symplectic chart leads to  Eqs. (\ref{waznya}), (\ref{waznyb}), which look the same for every Hamiltonian. Complete information concerning the potential is contained in symplectic connection coefficients so the eigenvalue equation for energy is covariant under any  change of potential. 
 
 To  write an explicit form of the Moyal product in the chart $(T, H)$
we  modified the Fedosov construction so that it was not necessary to  refer to the coordinates $(q,p).$

Among all the solutions to Eqs. (\ref{waznya}), (\ref{waznyb}) there are some unphysical ones. To eliminate them we propose two criteria: Theorem \ref{tw1} with its variant  (\ref{a6.1}) and Theorem \ref{tw1.2}. 
Unfortunately, the differential definition of the Moyal product is not really applicable to Wigner eigenfunctions, since they contain negative powers of $\hbar.$

The integral condition  (\ref{a6.1}) works only when coordinates are linear functions of $q$ and $p.$
Also Theorem \ref{tw1.2} has been  written in the coordinates $(q,p)$ and we were unable to find its covariant form. It may be that difficulties in expressing the formula (\ref{a6.1}) and Theorem \ref{tw1.2} in a form invariant under any canonical transformation are due to the fact that in quantum mechanics spatial and momenta coordinates are separable.

It seems to be possible to apply the proposed method of dealing with the eigenvalue equation to other observables and  phase spaces $({\mathbb R}^{2n}, \omega), \; n >1.$ We  are also considering the possibility of   generalizing the Theorems \ref{tw1} and \ref{tw1.2} to non-normalizable states. 

{\bf Acknowledgments}
 
This work was supported by the CONACYT (Mexico) grant No. 103478.

\end{document}